# Consciousness, Self and Language

Version 0.2

Robert Worden

Active Inference Institute, Davis, CA, USA

rpworden@me.com

**Abstract:**

Theories of consciousness depend on data, and it needs to be appropriate data, without overwhelming confounding factors. The reports of Minimal Phenomenal Experience (MPE) in [Metzinger 2024] relate to consciousness in a state purer than everyday consciousness, which may have fewer confounding factors.

This essay suggests that the confounding factors, which are absent or diminished in MPE states, are related to language. The self which is absent in mindful states is a product of language. The link between language and MPE states is demonstrated by reference to the phenomenal reports in [Metzinger 2024].

Language, emotion, and mindfulness are analysed in terms of Bayesian pattern matching, or equivalently minimisation of Free Energy, using three types of pattern which are specific to humans. These types are the word patterns of language, self-patterns which drive our emotions and which are also a part of language, and mindful patterns. The practice of mindfulness involves learning mindful patterns, which compete with self-patterns and displace them, allowing mindful states to occur. Consequences of this picture for theories of consciousness, and their relation to MPE states, are explored.

## 1. Introduction

In the science of consciousness, there has been an accumulation of theories, but no breakthroughs and no acknowledged leading theory. For instance, many theories of consciousness relate it to neural activity in the neocortex; but there is compelling evidence that there can be consciousness in the absence of part or all of the neocortex [Merker 2007]. We have little reason to place trust in most existing cognitive theories of consciousness; their sheer diversity reduces our trust in any one of them.

The science of consciousness today is like to the science of motion in the seventeenth century. Then the laws of motion lay hidden, because the available data were beset with confounding variables - notably, friction and obstacles. Studying a cart in a muddy field sheds little light on the underlying laws of motion. To discern the laws, we needed data where the confounding factors were diminished or absent. Isaac Newton found those data in the motion of the planets, which have no friction and no obstacles. Having found laws which accounted for planetary motion, it was possible to bring those laws back to earth, and to apply them in the presence of confounding variables.

The MPE project has investigated consciousness in states where certain confounding variables are reduced or absent, in a more pure state of consciousness which may be analogous to unhindered planetary motion. The project has produced a wealth of phenomenal reports, which could be instrumental in a better theory of consciousness. A first step towards such a theory is to ask: in what way are these meditative states 'pure' states of consciousness? What confounding variables are diminished or absent in MPE states?

This essay offers a possible answer. The confounding factors – found in everyday consciousness, but diminished or absent in MPE states – are the products of language. This is not just in a direct surface sense, that verbal thought is diminished or absent in mindful states, but in a broader sense, which can be traced across all the MPE reports in [Metzinger 2024]. Most important, the dual human self – which conflates an agent self (or ego) with the experiencer – is entirely a product of language. By their contrast with everyday consciousness, reports of non-dual states of consciousness reveal the influence of the linguistic agent self on conscious experience. Other factors reported as absent or diminished in MPE states – such as the will, desire, chronological time, and narrative self-history – are the products of language. This essay contains no new empirical data, but analyses the data in [Metzinger 2024] from the perspective of mindful states as consciousness without language. That perspective can illuminate some (but not all) that unifies the phenomenal reports.

The second part of the essay is more theoretical, asking what this can teach us about the cognitive basis of consciousness. We need to understand the whole influence of language in the human mind, resulting from its evolutionary origins. Language is a prodigious application of the Bayesian pattern-matching ability of the mind, devoted to a particular human purpose. I suggest that language is not just a neutral, transparent tool for describing reality (as many suppose it to be), but it is primarily a vehicle for discussing status within a social group. Thus language has created the human agent self, who chooses to act and thereby gains social status. Several other features of language are symptoms of its underlying status-related purpose.

We learn thousands of word patterns in our native language, and then do Bayesian matching of those patterns as we use language. This starts in early childhood; later, we never stop learning any word, and we never un-learn a word. Learning another large set of language-related patterns (which I call self-patterns) is the key to our emotions. In Eastern spiritual traditions, learning of self-patterns is called 'conditioning'. Learning a third kind of mindful pattern is a key to mindfulness. In this view, mindful patterns progressively displace our conditioned self-patterns, allowing mindful states to exist.

The resulting view is not an immediate key to unlocking a cognitive theory of consciousness; but by giving a grounded Bayesian account of certain purer states of consciousness, it may help us navigate past the confounding variables, towards such a theory.

## 2. Word Patterns and Self-Patterns

This section discusses the cognitive basis of human language and emotion, as a prelude to discussing MPE states of consciousness.

In recent years there has been a Bayesian revolution in cognitive science [Rao et al. 2002; Friston, Kilner & Harrison 2006; Friston 2010], with the realisation that the brain has prodigious Bayesian pattern-matching capability in higher animals and in mankind. The Bayesian brain hypothesis has been extraordinarily successful in fitting many types of empirical data on animal and human cognition [e.g. Krebs & Davies 1987, Anderson 1990, Parr, Pozzuolo & Friston 2022].

Two facts have been less widely appreciated in this Bayesian revolution. The first is the very large number of patterns which need to be matched against sense data, even in small animals. For instance, a bird or small mammal may need to recognise thousands of patterns for all the locations it visits in its habitat, to know what may happen in each location. The second fact is that nearly all these patterns need to be learned early in a short lifetime – requiring a prodigious capacity for fast learning of thousands of structured patterns, which we still cannot reproduce in neural models of the brain[1].

Both of these are evident in the human capacity for language. Our knowledge of our native language includes at least thousands of **word patterns**, for all the words we know and use. Each word pattern contains a sound part, matching the sound of the word, and a meaning part, often linked to conscious mental images of its meaning. There are now working computational models of how word patterns are matched[2], both in language production and understanding. All those thousands of word patterns need to be learnt, starting in early childhood.

The matching of word patterns in language is done very fast (matching several word patterns per second) and pre-consciously. We try out several different sets of word pattern matches in parallel, to find the Bayesian maximum likelihood meaning of what we hear (i.e. with the least Free Energy), before we are consciously aware of the most likely meaning. Bayesian word pattern matching is competitive, and we are not aware of the word patterns that lost in the competition.

Our language ability also includes the ability to carry on conversations – as studied in language pragmatics [Levinson 1983,2025; Sperber & Wilson 2012; Huang 2017]. Language pragmatics includes the ability to infer a conversational partner's knowledge, and their conversational intent (including the type of speech act), from what they say in the conversational context. This conversational mind-reading is done very fast [Levinson & Torreira 2012], so it is done by Bayesian pattern-matching. As well as word patterns, we learn large numbers of conversational **mind-reading patterns** as part of our native language. We constantly use them in conversation, without conscious awareness of them. Speakers and listeners use mind-reading patterns to build up a **common ground** of shared understanding [Stalnaker 2002, Tomasello 2014]. The common ground is fundamental to all language.

---

[1] Current neural nets can learn large numbers of patterns, but they learn them far to slowly to be useful to any animal.

[2] These models have been most fully worked out not in Chomskyan linguistics, but in the alternative approach of Cognitive Linguistics, or Construction Grammars [Hoffman & Trousdale 2013; Worden 2025a]. In construction grammars, words are a type of construction. A construction is represented by a feature structure, which is a directed acyclic graph (DAG). The central Bayesian pattern-matching operation is unification, which matches and merges feature structures, and is used to produce or understand any utterance. Language production and understanding have been implemented in computational models for many languages, at Marr's [1982] level 2, not as a neural implementation. There is no connection of this work to Large Language Models (LLM).

There is a subset of the mind-reading patterns which is important for our emotions. When we infer what a conversational partner is thinking, we infer what they are thinking about ourselves. As a simple example, when we ask somebody a question, we infer that they know we want an answer - that they know something about our own mental state. We also infer what they are thinking about our own social status – for instance: 'he will think I am being impolite.' or 'she will think I am ignorant.'. We may use these fast pre-conscious inferences to filter what we say in a conversation. We have many learned mind-reading patterns of the form 'If X, then person Y will think Z about my own social status' (X can be something I say, or do). As these patterns focus on our own social status, I call them **self-patterns**.

The human sense of self, which differs profoundly from the sense of self in all other animals[3], is derived from language and is learnt in language. It is primarily a sense of 'what I think other people think of me' (the self seen in the mirror of other peoples' thoughts) created by constant pre-conscious matching of self-patterns. I shall refer to this self as the **agent self** – the 'me' (or ego) who chooses to act, and whose social status is judged by other people, on the basis of my actions. The agent self is not a real biological entity; it is learned as a part of language. In language, it is a semantic 'agent' or grammatical subject, who chooses to act, in the active mood.

From childhood onwards, as part of our native language we learn thousands of self-patterns. As we speak to other people, or as we choose to do anything, or as we think privately[4], these self-patterns are constantly being matched, outside our conscious awareness. Together, they define our self-esteem, which drives our social emotions. In Eastern religious teachings, the learning of self-patterns is called conditioning. Mindfulness involves a release from conditioning. To understand mindful states, we need to understand how self-patterns are learned and used; and can they ever be un-learned?

Self-patterns have an important property. Groups of self-patterns can sustain themselves in a runaway cascade process. A first self-pattern lowers our self-esteem and causes us to feel some negative emotion; this is consciously experienced as a pattern of feelings in our body, which we may (or may not) label with a word such as 'shame'. The pattern of bodily feelings matches a second self-pattern, which may be paraphrased as 'if I feel this negative feeling, that is a low-status thing to do; if other people see me feeling this, I will have yet lower status' This in turn may trigger further self-patterns and negative emotional feelings, in a self-sustaining cascade process. This is the root of our volatile and irrational emotions; even without conscious thought, we may experience a mixture of bodily feelings which we associate with negative emotions. This may trigger conscious verbal thought, in an attempt to solve the problem. This pattern of background anxiety and discontent is part of our language heritage; Buddhists call it 'dukkha'. The self-sustaining cascade may be regarded as the Buddhist cycle of birth and re-birth; a continual re-birth of the agent self.

In this description of language and the human self as they are today, the reader may ask (from a view that 'nothing in biology makes sense, except in the light of evolution') how did language come to be that way? What selection pressures made it like this? Why is language biased towards issues of social status? There is a possible answer to these questions, in sexual selection [Miller 2001; Worden 2022, 2024] – which leads to strong species-specific selection pressures, strong enough to create our prodigious and unique language capability in a short evolutionary timescale. That answer is not central to this essay; interested readers may look at the references.

---

[3] For instance, most animals do not recognise themselves in a mirror. To them, life might be like a film show with no actor.
[4] In private verbal thought, when no other person is present, we imagine some 'shadow audience' who observes what we say and do, and assesses our social status from it. The same self-patterns operate with the shadow audience as with a real audience. In psychoanalysis, the shadow audience is called the Freudian Superego, or the 'parent' figure of Berne's Transactional Analysis [Berne 1964]

## 3. Analysis of MPE States

This section contains an analysis of the MPE states of consciousness reported in [Metzinger 2024], in terms of the hypothesis that mindful states are linked to diminished activity of word patterns and self-patterns.

Each chapter in [Metzinger 2024] contains many verbal reports of MPE states, usually from experienced meditators, often reporting experiences (inside or outside formal meditation) that they found remarkable in some way, rather than some more typical meditation experience. This is a selected sample of phenomenal reports. In scientific terms it can be regarded as data measured after careful preparation of the measuring apparatus, in the usual experimental fashion. Metzinger reiterates the risks of 'theory contamination' – a retrospective re-framing of the experience for verbal report, in the light of some theory of mindfulness.

The language hypothesis of mindfulness can be linked to the phenomenal reports in all chapters of the book. I take a selective approach: for selected chapters, where the links seem particularly clear or instructive, in the table which follows I interpret the phenomenal reports in terms of the language hypothesis – that mindful states are those in which the action of word patterns and self-patterns is diminished or absent.

After the table, I state a further hypothesis about how in mindful states, those patterns are diminished or absent. A key part of that hypothesis is that over time, we learn **mindful patterns** – in which some pattern of bodily feelings is matched not by any assessment of one's own social status (as in self-patterns), but by the internal action of paying closer attention to those bodily feelings, or to whatever else is happening - without interpreting it as having some meaning. A mindful pattern is a pattern of closer observation, without interpretation.

| Chapter | Interpretation in terms of Bayesian pattern matching |
|---|---|
| 1: Relaxation | When the cascades of self-patterns are reduced, the anxious background feeling and the associated bodily tension are reduced; this leads to feelings of relaxation. |
| 2: Peace | Reduction of self-pattern cascades leads to a changed balance of bodily feelings, and a reduced need to think out solutions to problems; both of which we experience as feelings of peace and reduced disturbance. |
| 3: Silence | Reduction of self-pattern cascades reduces the triggering of verbal thoughts, to try to solve some problem. When verbal thoughts occur, they are less likely to trigger further self-patterns; so the thoughts can be simply observed without the reaction of self-patterns. |
| 4: Wakefulness | The direction of causation here is not entirely clear. With the diminished action of self-patterns, some wakefulness is required, in order not to fall asleep; so wakefulness may precede the mindful state, rather than be caused by it. |
| 5: Clarity | When any conscious experience matches a self-pattern or word patterns, the mind may jump to classify the experience as 'that and only that', and make it fit the pattern, rather than explore the experience more fully. Matching mindful patterns involves closer exploration of any aspect of experience, leading to greater clarity of perception. |

| 8: Nonidentification | The agent self (or ego) is created by the matching of self-patterns, and the consequent triggering of word patterns; when this activity is reduced, what remains is conscious experience, without thought of a self experiencing it. |
|---|---|
| 9: Suchness | Suchness is related to clarity. When there is a reduced tendency to classify any experience as 'that and only that' – to make it fit some pattern – then each experience is experienced more deeply, and some pre-learned label no longer fully describes the experience. The extra experience beyond the labelled pattern is suchness (related to the Zen 'nothing special') |
| 10: Presence | Before language, all animal experience was experience of the present moment. Language introduced chronological time in its past and future tenses (possibly because they enable wider discussion of status-related issues).<br><br>With the reduced action of self-patterns, there is reduced action of word patterns for the future and past, and a timeless experience of being present. |
| 11: Connectedness | Self-patterns lead to the experience of a boundary between the physical space of the body and the rest of space, and to the thoughts 'this is me; that is other'. Reduced activity of self-patterns leads to a pure conscious experience of a three-dimensional space without a boundary between a self and other – a connected experience of limitless space, which contains all people and things.<br><br>This relates to compassion. When other people are no longer experienced as 'things out there that must be dealt with, to maintain my own status', the mind is freer to experience what life might actually be like for those people. |
| 12: The most natural state | This relates to the Zen 'nothing special' ; a combination of profundity and simplicity. Ordinary and unspectacular.<br><br>Language carves reality apart into 'special' things, for purposes of agency and distinguishing status. Without language, there is no unnatural division between the self and the rest of nature. |
| 13: Coming home | Home is a kind of ground state of the mind, in which there are no perturbations to be corrected. Matching of self-patterns is the source of perturbations, which need to be corrected by further self-patterns or by verbal thought. The absence of these perturbations is like an experience of coming home. |
| 14:There is nothing left to do | Language created the agent self or ego, which chooses to do things, and whose status is assessed by other people by the results of what it does. In choosing to act, the agent self exercises 'free will', which is an artefact of language.<br><br>When the agent self is diminished or absent, there is freedom from the will to do anything. This is the experience of nothing left to do. |
| 15: Joy, bliss and gratitude | Words like 'bliss' can be dangerous – implying that there is something to attain, and distracting from experiencing exactly what is. While bliss may occur, Tibetan Buddhism warns against striving for bliss.<br><br>On the other hand , it might be that when human beings evolved to be burdened with language and its self-patterns – producing , as they do, clusters of unpleasant feelings – our conscious experience had to be somehow compensated, by adding or enhancing some positive qualia of joy, love and gratitude, which are not experienced by other animals, in order to keep our average balance of feelings nearly neutral; to make us want to carry on. When evolution dealt us a negative card, it had to deal us something positive at the same time. These positive feelings are experienced when the self-patterns are absent. |
| 16: Simplicity, nothingness and absence | Here we move further into territory where retrospective verbal reports have no common ground to be based on, and so words may be ineffective.<br><br>Nevertheless, in the absence of verbal thought there is simplicity. For many things, their absence can be a positive experience. |
| 17: Emptiness and fullness | This paradoxical contrast reveals the limitations of language in describing these experiences. |
| | This chapter discusses the important topic of narrative self-deception, which is an intrinsic part of language. Language creates both the egoic agent self and the extended time of past and future. Between them, these concepts enable us to make up personal life stories, and to polish them in the service of our self-esteem. Since the life stories are personal to us, they are subject to little external criticism or checking; they may contain much that is questionable or false. Notably, the narrative is a story of an agent self which does not exist. |
| | For those interested in mindfulness, their personal life story includes their relation to mindfulness. Some aspects of their mindfulness stories may be unfounded – a narrative self-deception - yet still be useful tools in the practice of mindfulness. But in the MPE reports, as Metzinger says, narrative self-deception can create an unknown amount of theory contamination. |
| | A key element of every personal life story is a wish for it not to come to an end, and the resulting fear of death. |
| 18 Luminosity | A central part of language is spatial metaphor, which we use to make abstract concepts spatially imaginable, to consciously experience them [Lakoff & Johnson 1980]. Two metaphors for mindful experience are spaciousness and light. It is common to experience these in mindful states; so retrospective verbal reports of such states make frequent use of these metaphors. |
| | Metzinger relates luminosity to 'epistemic openness'. In the language interpretation, I find this hard to grasp; if 'epistemic' is related to knowledge, how does knowledge come from experience? What is knowledge without words? Is this an inevitable theory contamination? |
| 22 from timelessness to timeless change | Under chapter 10 on Presence, I discussed timelessness, as the lack of a linguistic tense which creates past and future; 'presence' is an unchanging present tense experience, as animals may experience it. |
| | Timeless change is another topic. It is possible that both animals and humans have qualia for change, which can be experienced at any location of their conscious space. For instance, spatial motion is a kind of change (precisely, a rate of change) and there may be motion qualia. Language has words for motion, because experiences of spatial motion can be shared with other people, in a common ground. But other kinds of qualia for change may not be shareable in words, as they are not part of any common ground. When our experience is not limited to what we describe in language, these kinds of 'timeless change' may be experienced. |
| 23 Space without centre, structure or periphery | In our normal spatial conscious experience, three-dimensional space is experienced with the middle of the head as an approximate centre of the space; and we identify the head as a centre of the agent self (it is what other people look at in conversation). In certain abnormal states of consciousness, notably in Out-of-Body Experiences (OBE), space is experienced with a different centre. So in states without an agent self (with no first-person perspective), we may experience space without a geometric centre. This may be reminiscent of Harding's [1964] 'On having no head'. |
| 24 Bodiless body experience. | Many meditators reported the disappearance of body boundaries. If body boundaries are the experienced consequences of a linguistic agent self, then, as in chapter 11 (connectedness), in the absence of an agent self, spatial consciousness may include no experience of a boundary between self and other. |
| | The absence of an agent self/ego relates to the absence of a Unit of Identification (UOI) for conscious experience. |
| 25: Ego dissolution: melting into the phenomenal field | In this chapter, the experience of an egoic agent self (and its disappearance) are directly related to the disappearance of an Epistemic Agent (EA). The EA, or agent self, is a product of language, which is not experienced when linguistic word patterns are inactive. |
| | Metzinger notes the 'myth of cognitive agency' that 'the paradigmatic case of conscious cognition is one of autonomous, self-controlled rational thought' – and then notes "It isn't". The rationally choosing agent self is a creation of language – created only to support the linguistic concept of choice, which leads to altered social status [Worden 2024] |

| 26: Nondual being: unity | In the grammatical subject (= semantic agent), specifically in the subject 'I', language has conflated two separate concepts: (1) the agent self who chooses to act, whose social status is judged by the results, and (2) the experiencer, which experiences whatever happens to it. This duality is built into the heart of language; we only escape from it to nondual states when the self-patterns of language are inactive. |
|---|---|
| | As well as contemporary phenomenal reports, Metzinger cites many historic writings, including Christian mystics such as Meister Eckhart [Walshe 1979]. These writings perhaps show that release from the dual agent self can occur within a wide variety of intellectual frameworks. Conceiving oneself as only a 'feather on the breath of God' can be a path to release from the self, and may be the spiritual core of all theistic religions. |
| 27: Nondual awareness: insight | The role of language and the absence of self-patterns relate directly to non-dual awareness, as above; but insight may be more problematic. The problem may relate to the cloud of meanings of the word 'insight'. Does it overlap with the meaning cloud of the word 'knowledge'; and to that extent, do insight and knowledge require words to express them? |
| | There is no doubt that meditators have some experiences in meditation which involve a release from language-based categories, and an absence of language-based thought; and that tese experiences are highly valued, and may be retrospectively described as 'insight'. But whatever insight is, it is not part of the common ground where language can operate; so perhaps we are best advised to remain silent, like Wittgenstein. |
| | There is a notable prevalence of birdsong in the phenomenal reports. Perhaps this dates back to something pre-human in the forest. |
| 28: Transparency, translucency, and virtuality | Spatial consciousness is consciousness of a Bayesian maximum likelihood model of the space around us, constructed by pattern-matching of sense data of all types; this is perhaps the most prodigious and important Bayesian pattern-matching done by any brain. Yet this model of space is obviously not reality itself; it is a virtual model, created by the brain – in some sense, a controlled hallucination. |
| | Yet for most of the day, the virtuality is not evident to us; the experience is transparent or translucent, as if we were experiencing real space, rather than a model of space. |
| | The phenomenal reports of MPE are that objects in the experienced space have an 'as if' quality, like objects in a virtual reality. This is an experience of the model-ness of things; the virtuality is no longer transparent. Such experience has some relation to the absence of self; the insight that the self is not real spreads out to an insight that other things are not real. Here I have used the problematic word 'insight' for non-verbal knowledge, because it seems to fit. |

More could have been said here, about other chapters in the book and about the many links between chapters. I hope I have said enough to support the view that many aspects of minimal phenomenal experiences can be usefully understood as the absence of language in its matching of word patterns and self-patterns. There follows a hypothesis of how mindfulness comes about.

## 4. How Mindful States Happen

In what follows, I assume that all higher animals have conscious experience, experiencing a conscious model of the three-dimensional space around them, including all the things they can sense (see, or hear, or feel) in that space. I assume that this conscious model is continually maintained by fast Bayesian pattern-matching of all types of sense data, with strong Bayesian priors about fundamental facts such as the Euclidean geometry of space, the linear nature of free motion, and the vertical pull of gravity[5]. This prodigious Bayesian pattern matching (or joint minimisation of Free Energy) creates the information content of all consciousness, including our own. Scientists are still in the process of discovering how the brain creates this information (the generative model of reality, which generates sense data). Please assume that as background for what follows.

---

[5] Having been true for all time, these priors have had plenty of time to evolve into brains

In the human brain there are three further types of pattern, all involved in the fast Bayesian pattern matching which precedes conscious experience:

1. **Word patterns** of language, which we use to speak, understand, and think
2. **Self-patterns** of language, which we use to infer how other people regard us as social beings.
3. **Mindful patterns**, in which some sense input leads us to explore it more deeply

In each one of us, all three types of pattern can be active at any time of the day. Word patterns are needed to speak, listen and think; we use self-patterns to define our role in society, and they are essential for all the functioning of society; and finally, we may have mindful moments at any time of the day – whether from music, dance, sporting activity, absorption in some task, a piece of nature, a work of art, or some Proustian involuntary memory triggered by a flavour or a sound. The problem is that for most of us, the mindful moments are just short moments, and are soon overtaken by self-patterns.

Once having learnt a word pattern, we never un-learn it. As self-patterns are parts of language, the same applies to them. If we can never un-learn any self-pattern, how can we escape from self-patterns? Fortunately there is a way.

Bayesian pattern matching is competitive; the pattern which best fits the evidence wins the competition, and is the only one that comes to conscious experience. We are never aware of any pattern that loses the competition; it is 'as if it had never been born'. We can see this in language, in ambiguous utterances such as 'recognise speech', which could also be heard as 'wreck a nice beach'. A child may learn the words 'wreck', 'nice' and 'beach' and never un-learn them; yet later, having learnt the words 'recognise' and 'speech', she can correctly find that meaning, when it is the most likely meaning in the context. That is how Bayesian pattern matching handles the ambiguities of everyday language.

Self-patterns match a pattern of feelings in the body. It is usually not a pattern of intrinsically unpleasant feelings, but it may be a pattern with several parts, that we do not usually describe in words. If we tried to describe one such pattern, it might for instance be 'a fuzzy feeling behind the eyes, together with a kind of 'gulp' feeling in the throat, and a hunching of the shoulders'. Whatever the pattern of real body feelings, whenever that pattern matches with a self-pattern, that self-pattern is triggered, leading to its consequences: 'As I am feeling this, people might think I have low status; I'd better cover up, or think of some way out…'. These are the cascades of emotion. There may be thousands of self-patterns, each having a different configuration of feelings in the body.

The practice of mindfulness is a practice of closely observing feelings in the body – the same feelings which are caused by self-patterns, and which match further self-patterns. The key to the practice is to observe closely; when you experience some piece of feeling, go deeper into it and experience it in finer detail. The aim is to experience the feeling exactly as it is – not some interpretation, or decoding, or meaning of the feeling. As you experience the more detailed pattern, you inevitably learn a new pattern. The fundamental process of mindfulness is to learn mindful patterns – patterns of detailed bodily feelings, whose only consequence is a finer, more detailed experiencing of those feelings.

If, during mindful practice, you experience the feeling resulting from some self-pattern, and matching another self-pattern, you can go on to experience it more deeply – thereby learning a mindful pattern, which is more detailed than the matched self-pattern. In future, the mindful pattern will compete with the self-pattern; and being more detailed, it will win the competition. You will no longer experience consequences of the self-pattern. Over time, you can do this 'replacement learning' for most of your many self-patterns – learning many mindful patterns, so that self-patterns almost never win the competition. You can remain in mindful states not just for moments, but for longer parts of the day.

Self-patterns decode some configuration of bodily feelings, to a meaning which leads to lowered social status for the agent self (in the eyes of some shadow audience). In mindful patterns, there is no decoding - just a closer experiencing of bodily feelings.

I have described three types of pattern in the human mind, and how they influence our conscious experience. The relative preponderance of those patterns in various states can be shown in diagrams – where the green bars represent mindful patterns, the blue bars represent word patterns, and the red bars represent self-patterns:

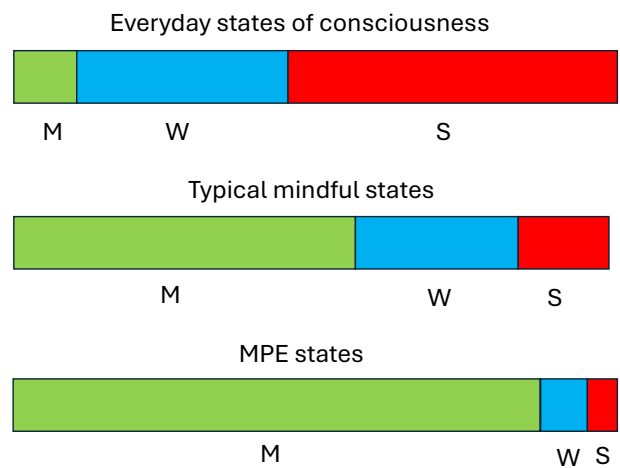

*Figure 1: Relative preponderance of three kinds of pattern matching, in three states of consciousness. M = Mindful Patterns; W = word patterns; S = self-patterns*

In everyday consciousness, mindful patterns are occasional and momentary. Self-patterns generally outnumber word patterns, because word patterns may trigger a cascade of many self-patterns. In typical mindful states, mindful patterns compete with self-patterns, and win often enough to diminish or remove the cascades of self-patterns. Word patterns may not be so much reduced, but because they do not create so many cascades of self-patterns, they dominate over self-patterns. Verbal thoughts can then be observed with little reaction. In MPE states, there is very little language (very few word patterns) and little or no self, or ego.

This spectrum does not capture the full variety of mindful states. For instance, a few people (such as Meister Eckhart, or J. Krishnamurti) have carried on active lives (talking, listening and acting), while not experiencing anything of the agent self, or ego. Their word patterns and other patterns are active, without a trace of self-patterns. This may also happen as a result of Zen training [Watts 1957], and does not fit into a single spectrum of mindful states. Conversely, the lives of some powerful people are dominated by self-patterns; they can be pitied, rather than envied.

In summary, the essence of mindfulness is a learning process  - learning which happens inevitably when you pay close attention to your bodily feelings just as they are, not as you would wish them to be, or as you interpret them. After some of this learning process, the mindful patterns which you learn will successfully compete with your old self-patterns, effectively displacing them from your experience. Mindfulness then happens without trying, by a pre-conscious Bayesian pattern competition.

## 5. Towards a Theory of Consciousness

Science depends on experimental data; if we are to find a theory of consciousness which is an advance on today's theories, it may depend on new types of data. The consciousness experienced in meditative states, including a minimal phenomenal experience (MPE), has great promise in this respect; it offers a new window into consciousness in a pure state. Reports of MPE may be analogous to the data on planetary orbits, summarised in Kepler's laws, which led to the first modern theory of mechanical motion.

To make best use of data, we need a theory of the measuring apparatus, and an (overlapping) theory of confounding factors in the data. In this case, asking for a theory may be asking too much; but we at least need hypotheses. For the planetary orbits, one hypothesis about the measurement process was that light from the planets travels in straight lines, and travels very fast; this hypothesis might have been wrong, but in the event it served[6].

This essay has described a hypothesis about the measuring apparatus and confounding variables for states of consciousness, including MPE states. The base hypothesis about the measuring apparatus is that the engine providing the data for the controlled hallucination of consciousness is the Bayesian brain, using sense data of all modalities. This part of the hypothesis is uncontroversial, because the Bayesian brain hypothesis is well established. The Bayesian brain provides data for consciousness, but it is hard to see how it creates consciousness – both because of the Hard Problem of consciousness [Chalmers 1996], and because all information in a computer brain is encoded in neural spike trains, yet consciousness contains un-encoded information about external reality, such as the shapes of boundaries in space [Worden 2025b]; the information required to decode neural spike trains does not reside inside the brain. I pass by these problems, the second of which is not widely appreciated, and move to other confounding factors in the data.

I have proposed that everyday consciousness is made complex and hard to understand by two kinds of Bayesian pattern matching which are both unique to humans – matching of language word patterns, and matching of the closely related self-patterns. This is the hypothesis of the confounding factors – that MPE states of consciousness differ from everyday states by the absence or near-absence of these two kinds of pattern-matching. States of consciousness can be placed on a spectrum depending on the prevalence of these patterns; or equivalently, on a level of mindfulness.

I have gone into some detail of a proposed natural history of mindfulness, to start to characterise mindful states; but much remains to be explored. I mention only a few possible directions:

1. On the assumption that higher animals are conscious, the mechanism of consciousness is older than language, including its self-patterns and the agent self. So self-consciousness is then not a fundamental part of consciousness; it is something which arrived with language,

---

[6] If light travelled more slowly, it would have made Kepler's laws more complex, and the discovery of the laws of motion more difficult

after consciousness existed. In this case, theories of consciousness which place self-consciousness at centre stage may not be favoured, unless we assume that animals are not conscious, and that all consciousness started with humans.
2. Verbal reports of MPE states have an unknown amount of theory contamination. Perhaps if we have a computational model of language and the self, including its creation of chronological time through tenses, and its creation of narrative life histories, we can develop a better understanding of narrative self-deception and theory contamination.
3. If MPE states are at the end of a spectrum of mindful states, where in pseudo-mathematical terms, language = 0, then verbal reports of MPE states are in some sense the first derivative of language at the point language = 0. Could this pseudo-mathematical sketch be developed into a respectable analysis?
4. In most MPE reports, conscious experience includes spatial experience – supporting the hypothesis that a base level of consciousness includes spatial consciousness. However, in a few reports of MPE states, there is not even space. How are we to interpret those reports? Are they the very end of a spectrum of mindful states, or might they be a result of theory contamination?
5. Some writings on mindfulness, such as the work of J. Krishnamurti [1970], can be seen to be consistent with the hypotheses of this paper, but only by carefully analysing his use of words; for instance, he uses the words 'Intelligence' and 'Intellect' with almost opposite meanings (intelligence is mindful, intellect is not). Can verbal reports of MPEs be clarified by exploring with the authors of those reports how they use words?
6. There are open questions about the relation of the Bayesian view of language (and of its absence, in MPE states) to the Free Energy Principle, and Active Inference [Parr, Pezzuolo & Friston 2022]. What kinds of free energy are represented by word patterns, self-patterns and mindful patterns? For instance, is the social status evaluation of self-patterns a part of Variational Free Energy (VFE), or of the Expected Free Energy (EFE) of Active Inference - which involves tradeoffs between exploitation of present value, and exploration for epistemic value? Epistemic value is the value of information in planning for the future ('expected' free energy), but the future is not relevant to MPE states. Has epistemic value any relation to the 'epistemic openness' of MPE reports? These are difficult questions, and I do not have answers.

I hope readers are stimulated to explore these and other directions for themselves.